\documentclass[prl,superscriptaddress,twocolumn,amsmath]{revtex4}

\usepackage{epsf,latexsym}
\usepackage{amssymb,amsmath}
\usepackage{times,graphics}
\usepackage{color}
\usepackage{epstopdf}
\usepackage{epsfig,ulem}
\newcommand{\ket}[1]{| #1 \rangle}

\def\6{\langle}
\def\9{\rangle}

\newcommand{\etal}{{  \textit{et al.}}}

\newcommand\p{{\sf p}}
\newcommand\w{{\sf w}}

\def\half{\tfrac{1}{2}}

\newcommand{\ignore}[1]{}

\newcommand{\be}{\begin{equation}}
\newcommand{\ee}{\end{equation}}
\newcommand{\ba}{\begin{eqnarray}}
\newcommand{\ea}{\end{eqnarray}}

\begin{document}

\title{Determinism, Independence and Objectivity are Incompatible}

\author{Radu Ionicioiu}
\affiliation{Department of Theoretical Physics, National Institute of Physics and Nuclear Engineering, 077125 Bucharest--M\u agurele, Romania}
\affiliation{Research Center for Spatial Information -- CEOSpaceTech, University Politehnica of Bucharest, 313 Splaiul Independen\c tei, 061071 Bucharest, Romania}

\author{Robert B. Mann}
\affiliation{Department of Physics and Astronomy, University of Waterloo, Waterloo, Ontario, Canada}
\affiliation{Perimeter Institute for Theoretical Physics, Waterloo, Ontario, Canada}

\author{Daniel R. Terno}
\affiliation{Department of Physics and Astronomy, Macquarie University, Sydney, NSW,  Australia}
\email{daniel.terno@mq.edu.au}
\begin{abstract}
Hidden-variable models aim to reproduce the results of quantum theory and to satisfy our classical intuition. Their refutation is  usually based on deriving predictions that are different from those of quantum mechanics. Here instead we study the mutual {compatibility} of apparently reasonable classical assumptions. We analyse a version of the delayed-choice experiment which ostensibly combines determinism, independence of hidden variables on the conducted experiments, and wave-particle objectivity (the assertion that quantum systems are, at any moment, either particles or waves, but not both). These three ideas are incompatible with any theory, not only with quantum mechanics.
\end{abstract}
\maketitle

\noindent \textit{Introduction.}
 --- Most of the quantum formalism was in place by 1932 \cite{vN}. Since then, quantum theory has been spectacularly successful across all the investigated scales and systems. Yet many of its results  contradict both  common sense and classical physical intuition. Wave-particle duality, superposition, and entanglement are among these counterintuitive features \cite{qf1, qf2} and the ``strictly instrumentalist" \cite{peres} core of quantum theory abandons many familiar traits of classical physics. As a result, there are profound differences of opinion on the meaning of quantum theory and the desire to explain or even to remove its puzzling properties \cite{qf1,qf2,peres,bell}.

Hidden-variable (HV) theories endeavor to give a satisfactory representation of our intuition while reproducing the experimental predictions of quantum theory \cite{qf1,qf2,peres,bell,branyan}. Imposing classical concepts (determinism, versions of locality, etc) on HV models constrains the resulting probability distributions. This may lead to ``paradoxes," i.e., an incompatibility of the allegedly reasonable assumptions with the predictions of quantum theory.

With the advent of quantum technologies \cite{qinfo, qtech} we can now realize classic \textit{gedankenexperiments} and develop new tests to confront the predictions of HV theories with those of quantum mechanics. When the latter are experimentally confirmed, HV models fail the crucial test of adequacy and, unless some loophole for the experiment is found \cite{santos}, should either be abandoned or amended to include deep, possibly unacceptable \cite{bell}, conspiratorial correlations. The loopholes, in turn, may be countered by more sophisticated set-ups \cite{overloop}.

Implicit in these debates is the premise that classically reasonable assumptions form a world view which, although experimentally inadequate, is nevertheless consistent. We question this tacit assumption and investigate the mutual compatibility of three classical requirements (determinism, independence, objectivity). Specifically, in the context of wave-particle duality and delayed-choice experiments \cite{wdc1,rmp} we inquire if it is possible to find \textit{any} probability distribution that satisfies all  three classical constraints. Here we answer this question in the negative: determinism, independence, and objectivity are incompatible; i.e., no such probability distribution exists.

Our work is motivated by two recent developments in quantum foundations. The first is the attempt to reconstruct quantum mechanics (QM) starting from physically motivated (operational, information theoretical) axioms \cite{hardy, ls, chiribella, fuchs}.

The second development brought a change in our understanding of the wave function \cite{pbr, cr2}. Specifically, the authors of \cite{cr2} show that ``[...] a system's wave function is in one-to-one correspondence with its elements of reality.'' This suggests that we have to reconsider long-standing views regarding the use of classical concepts as obsolete in QM.
Before studying the tensions between QM and the classical world view, it is worthwhile to investigate the latter's internal consistency. Wave-particle objectivity and its generalization are seemingly milder requirements that are perfect candidates for such a study.

The context in which our results are relevant is twofold. First, as an attempt to understand which of the classical intuitions fail, what are the relationships between them, and which set of assumptions are mutually compatible or not. We show that although any two of the three assumptions are compatible, all three are not. This puts restrictions on future reconstructions/developments in quantum foundations. And second, as an attempt to identify which classical ``elements of reality'' can be preserved and which will have to be rejected.  The method that we use is based on an extension of the HV formalism to describe wave-particle duality \cite{wdcq}.

\begin{figure}[tbp]
\includegraphics[width=0.48\textwidth]{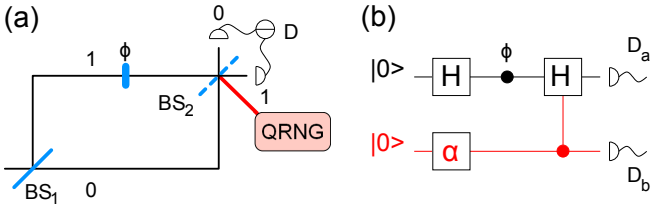}
\caption{The evolution of the delayed-choice experiment. (a) In Wheeler's classic experiment, the second beam-splitter is inserted or removed after the photon is inside the interferometer. The detectors observe either an interference pattern depending on the phase $\varphi$ (wave behaviour), or a flat (constant) distribution of hits (particle behaviour) \cite{wdc1}. A quantum random number generator (QRNG) determines whether $BS_2$ is inserted or not.  (b) Quantum network. The beam splitter is equivalent to a Hadamard gate \cite{qinfo}. The QRNG is an auxiliary quantum system initially prepared in the equal superposition state $\ket{+}= \tfrac{1}{\sqrt{2}}(\ket{0}+ \ket{1})$ and then measured. In the delayed-choice experiment with a quantum control the Hadamard gate is controlled by the ancilla prepared in the state $\cos\alpha|0\9+\sin\alpha|1\9$ and can be measured after the photon is detected by $D_a$ \cite{wdc2}.}
\label{fig1}
\vspace{-5mm}
\end{figure}

\noindent \textit{Setting.}
--- Wave-particle duality and the opposing idea of wave-particle objectivity are best illustrated in the  Wheeler delayed-choice experiment  \cite{wdc1,rmp}, Fig.~\ref{fig1}. We briefly introduce it and then proceed to our generalized model.

From a classical perspective, quantum systems (like photons, electrons, etc.) behave in complementary ways -- what one would call ``waves'' or ``particles.'' A single-photon interference (a definite wave-like behaviour), is produced by particle-like single-photon detections \cite{gra86,wdc1,wdcq,wdc2}. Hence we adapt as operational definitions of ``wave" and ``particle" counting statistics  { of individual detectors} as {dependence (independence) on} the phase shift {in} one of the arms of the Mach-Zehnder interferometer (MZI).  Wave data are obtained when the MZI is closed (second beam splitter inserted); particle data when it is open (second beam splitter absent), providing us with the operational definitions of the wave statistics
\be
e_\w=\left(\cos^2\frac{\varphi}{2},\sin^2\frac{\varphi}{2}\right), \label{dw}
\ee
and, respectively, particle statistics
\be
e_\p=\big(\half,\half\big), \label{dp}
\ee
We assume that the beam splitters are balanced (50/50) and polarization insensitive.

The complementarity \cite{qf1,qf2,bohr} of setups \cite{gra86} of the MZI needed to observe the particle and the wave behaviours allows one to entertain an  objective  view \cite{wdc2,six} that at any moment of time a photon is \textit{either} a particle \textit{or} a wave. Randomly choosing whether or not to insert the second beam splitter after the photon enters the interferometer (Fig.~\ref{fig1}a) prevents a possible causal link between the experimental setup and the photon's behaviour \cite{wdc1,rmp,six}.

In the quantum delayed-choice experiment (Fig.~\ref{fig1}b) one can first detect the photon and only later find out the type of test performed \cite{six,qdc-exp1,qdc-exp2}. In this experiment we need a variable bias $\alpha$ in order to observe the morphing behaviour from particle-like into wave-like statistics \cite{wdc2}.

In quantum mechanics the joint state of the photon $A$ and ancilla $B$ just before the measurements is:
\be
|\psi\9= \cos\alpha|\p\9|0\9+ \sin\alpha|\w\9|1\9,
\ee
where the wavefunctions $\ket{{\sf p}}= \tfrac{1}{\sqrt{2}} (\ket{0}+ e^{i\varphi}\ket{1})$ and $\ket{{\sf w }}= e^{i\varphi/2}(\cos\tfrac{\varphi}{2}\ket{0}- i \sin\tfrac{\varphi}{2}\ket{1})$ result in particle and wave statistics, respectively \cite{wdc2,six,qdc-exp1,qdc-exp2}. We represent the counting statistics as a vector of relative frequencies and arrange the entries alphanumerically, $ab=00,01,10,11$. With this notation, the statistics predicted by quantum theory is:
\be
\label{qmpredict}
q(a,b)= \big(\half \cos^2\!\alpha, \sin^2\!\alpha \cos^2\!\tfrac{\varphi}{2}, \half \cos^2\!\alpha, \sin^2\!\alpha \sin^2\!\tfrac{\varphi}{2} \big).
\ee

\begin{figure}[tbp]
\includegraphics[width=0.28\textwidth]{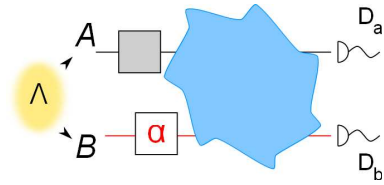}
\caption{An abstract rendering of the delayed-choice experiments. Two distinct statistics for the system $A$ are observed depending on the setting of the ancilla $B$. A hidden variable theory assumes that the state of $A$ and $B$ is fully determined by $\Lambda$.}
\label{fig2}
\end{figure}

We now introduce an abstract setting (see Fig.~\ref{fig2}) which separates the classical assumptions (leading to the delayed-choice ``paradoxes") from quantum mechanics. While we use the network language to describe the evolution of the system, we do not make any specific quantum-mechanical assumption regarding it. As before, the type of measurement on the system $A$ is determined by the setting of the ancilla $B$ which is revealed by the outcome $b=0,1$ of the detector $D_b$. Two statistically distinguishable probability distributions are observed for the system $A$,
\begin{eqnarray}
\bar{e}_\p(a) &\equiv& e(a|b=0)= (e_p,1-e_p), \nonumber \\
\bar{e}_\w(a) &\equiv& e(a|b=1)= (e_w,1-e_w), \label{emp12}
\end{eqnarray}
for some numbers $0\leq e_p,e_w\leq 1$, where $\bar{e}_\p(a)$ represents the ``p-statistics" (analogously to the particle-like behaviour in the open MZI), and $\bar{e}_\w(a)$ represents the ``w-statistics" (analogously to the wave-like behaviour in the closed MZI). {We assume that we can freely choose a parameter $\alpha$ on system $B$ and that the statistics of $D_b$ depends on this choice:}
\be
e(b)= (x, 1- x), \label{xal}
\ee
with $0\leq x(\alpha)\leq 1$. This is more general than the quantum delayed-choice experiment --- any experimental setup in which the output statistics of $D_b$ depend on $\alpha$ is sufficient for our purpose. In the following, we use the standard rules for the marginal $p(i)= \sum_j p(i,j)$ and conditional probability distribution $p(i,j)= p(i|j) p(j)= p(j|i) p(i)$ (Bayes' rule).

From the two conditional probability distributions \eqref{emp12} and the marginal statistics $e(b)$ we can reconstruct the joint distribution $e(a,b)$:
\be
e(a,b)= \big(x e_p, (1-x)e_w, x(1-e_p), (1-x)(1-e_w) \big).
\label{general}
\ee
Any general probability distribution $e(a,b)$ can be represented by three independent parameters $x$, $e_p$, $e_w$. For the quantum delayed-choice experiment {these parameters} are defined by Eqs.~\eqref{dw}, \eqref{dp} and \eqref{qmpredict}.

\noindent \textit{Hidden-variable model.}
--- A HV theory is encapsulated in two elements: a conditional probability distribution of the observable quantities given the value of HV $\Lambda$, $p(a,b,\ldots|\Lambda)$ and a probability distribution of $\Lambda$, $p(\Lambda)$. The observed probabilities are obtained by an appropriate integration or summation.

A HV theory is adequate \cite{peres, qf2, branyan} if it reproduces the  experimentally observed statistics. Here we investigate not if a proposed HV theory is adequate (in a world described by quantum theory), but if any statistics with marginal distributions \eqref{emp12} and \eqref{general} can be based on it. To establish our claim we prove that no probability distribution
\be
e(a,b)= p(a,b)= \sum_{\Lambda} p(a, b,\Lambda)= \sum_{ \Lambda} p(a, b| \Lambda)\, p(\Lambda),
\label{qp}
\ee
exists, without assuming anything about the parameters $x$, $e_p$, and $e_w$ apart from a generic dependence of $x$ on the settings $\alpha$.

HV theories intend to complete or improve quantum mechanics by incorporating classical intuitions, and thus should satisfy additional properties. We consider the reification of the counting statistics of Eq.~\eqref{emp12} summarised by the following:

(i) \textit{Objectivity}. We objectify the statistics given by $\bar{e}_\w$ and $\bar{e}_\p$ as {reflecting an} intrinsic property of the system, like ``wave" or ``particle" in the delayed choice experiment, which is unchanged during its lifetime \cite{wdc2,six}. This property is expressed by a binary function $\lambda=\w,\p$  of the HV $\Lambda$, $\lambda=\lambda(\Lambda)$. This is a property of an individual system, but could be causally influenced by changing the experimental settings. It is revealed in one setting of the apparatus (e.g., in the closed MZI) as
\be
p(a|b=1,\lambda=\w)=\bar{e}_\w(a),
\ee
and in another setting as
\be
p(a|b=0,\lambda=\p)=\bar{e}_\p(a).
\ee

(ii) {\em Determinism}: a knowledge of the hidden variables $\Lambda$ determines the individual outcomes of the detectors. This is a standard feature of HV models \cite{vN,qf1,qf2,peres}. We demand its weak form \cite{branyan}
\be
p(a,b|\Lambda)= \chi_{ab}(\Lambda), \label{cf2}
\ee
where the indicator function $\chi=1$, if $\Lambda$ belongs to some pre-determined set \cite{branyan}, and $\chi=0$ otherwise.

(iii) {\em Independence}: the property of $\lambda$-independence \cite{branyan,six} assumes the nature of the system, as determined by the value of a hidden variable, does not depend on the experimental setting. In our context it means that choosing $\alpha$ [in the quantum delayed-choice experiment, the rotation $R(\alpha)$] does not affect $\Lambda$. In line with the standard HV practice we assume the setting $\alpha$ can be selected independently \cite{qf1,qf2,peres,bell, branyan}. To exclude the causal influence permitted by (i) and to enforce (iii), experiments ensure a spacelike separation between parts of the system and assume absence of the superluminal communications \cite{qf1,qf2,peres,bell,wdc1,rmp}.

\noindent \textit{Solution to the constraints.}
Now we show that there is a unique non-trivial assignment of the probabilities $p(a,b,\lambda)$. The eight probabilities $p(a,b,\lambda)$ are normalised and adequacy conditions can be written as
\be
e(a,b)= p(a,b)= p(a,b,\p)+ p(a,b,\w).
\label{adeabc2}
\ee

In addition, (i) and the standard rules for the conditional probabilities, such as
\be
p(a|b, \lambda)= \frac{p(a,b,\lambda)} {p(0,b, \lambda)+ p(1,b, \lambda)}
\ee
imply two additional constraints,
\begin{align}
& p(0, 0, \p)(1- e_p)= p(1, 0, \p)\, e_p, \\
& p(0, 1, \w)(1- e_w)= p(1, 1, \w)\, e_w. \label{drealism-p2}
\end{align}

The resulting linear system has a two-parameter family of solutions $p_2(a,b,\lambda)$. However, for the generic solution with $p(b=0,\lambda=\w)\ne 0$, $p(b=1, \lambda=\p) \ne 0$ the resulting statistics is independent of $\lambda$,
\begin{align}
p_2(a|b=0,\p)&= p_2(a|b=0,\w) =\bar{e}_\p(a), \\
p_2(a|b=1,\w)&=p_2(a|b=1,\p) =\bar{e}_\w(a),
\end{align}
i.e., the statistics of $D_a$ is determined solely by the state of the apparatus. Any such theory reintroduces w-p  duality (analogously to the wave-particle duality in the delayed choice experiments) and therefore, nullifies objectivity assumption (i).

We can construct a nontrivial HV theory using a special solution
\be
p_s(b| \lambda)= \delta_{\lambda \p}\delta_{b0} +\delta_{\lambda \w}\delta_{b1}\equiv p_s(\lambda| b)
\label{blambda2}
\ee
which introduces a perfect correlation between $b$ and $\lambda$.

\noindent \textit{The contradiction.} 
---We now show that accepting Eq.~\eqref{blambda2} as a way to preserve the HV model contradicts the assumption (iii). From Eqs. \eqref{adeabc2} and \eqref{blambda2} we have
\be
p_s(\lambda)= \sum_b p_s(\lambda|b) p(b)= \sum_b(\delta_{\lambda\p}\delta_{b0}+\delta_{\lambda\w}\delta_{b1}) e(b),
\ee
This immediately implies that $\lambda$ has the same statistics as $b$, as expected from \eqref{blambda2}
\be
p_s(\lambda)= (x(\alpha), 1-x(\alpha)) \equiv p[\lambda(\Lambda)],
\ee
where the last equality holds as a result of (i). The left-hand side depends on $\alpha$, but the right-hand side does not, since $\lambda(\Lambda)$ is some fixed function of the HV which is independent of the parameter $\alpha$, as required by (iii). This proves the contradiction.

\noindent \textit{Discussion.} This result implies the following. The assumptions (i)-(iii) are consistent in classical physics, where all systems behave either as particles or as waves. However, if the same system (e.g., a photon) demonstrates two types of statistics (particle or wave statistics in the delayed choice experiments) in two different experimental setups (MZI open or closed), then it is impossible to construct a causal deterministic theory which promotes the two observed statistics to the status of objective properties of the system.

We stress that these statistics do not need to be derived from the quantum predictions [such as Eq. \eqref{qmpredict}] --- all that is required is that different setups yield different statistics.  {We  are also not imposing any constraints on possible correlations between several systems}. Consequently, this result does not depend on comparison of the predictions of a candidate HV theory with quantum mechanics (compare with \cite{pt02}).

Any two constraints are mutually compatible in our setting. It seems natural to drop the objectivity (i) from the list of classical desiderata. It is known that weak determinism and $\lambda$-independence [constraints (ii) and (iii)] are consistent with quantum mechanics \cite{branyan}. Nevertheless, before an attempt to supplement quantum mechanics can start, one of its counterintuitive  features (say, wave-particle duality) must be accepted. Whether this indicates a failure of the HV program or not is a matter of opinion. Our work establishes that there are situations where plausible classical ideas are not mutually compatible.

\medskip

\noindent \textit{Acknowledgments.} It is a pleasure to thank Lucas C\'{e}leri, Mile Gu, Peter Knight, Stojan Rebi\'c, and Vlatko Vedral for useful discussions, and the Centre for Quantum Technologies of the National University of Singapore for hospitality. This work was supported in part by the Natural Sciences and Engineering Research Council of Canada.


\begin{thebibliography}{99}

\bibitem{vN} J. von Neumann, \textit{Mathematische Grundlagen der Quantenmechanik}, (Springer, Berlin, 1932); transl. by E. T. Beyer, \textit{Mathematical Foundations of Quantum Mechanics}, (Princeton University Press, 1955).

\bibitem{qf1} J.A.~ Wheeler and W.H.~Zurek, (eds.), \textit{Quantum Theory and Measurement}, (Princeton Univ.~Press, 1984).

\bibitem{qf2} D. Greenberger, K. Hentschel, F. Weinert,  (eds.), \textit{Compendium of Quantum Physics}, (Springer, Berlin, 2009).

\bibitem{peres} A. Peres, \textit{Quantum Theory: Concepts and Methods} (Kluwer, 1995).

\bibitem{bell} J.S. Bell, {\it Speakable and Unspeakable in Quantum Mechanics} (Cambridge Univ. Press, Cambridge, 1987).

\bibitem{branyan} A.~Brandenburger and N.A.~Yanofsky, J. Phys. A {\bf 41}, 425302 (2008).

\bibitem{qinfo} M.A.~Nielsen and I.L.~Chuang, \textit{Quantum Computation and Quantum Information} (Cambridge Univ. Press, 2000); D. Bru{\ss} and G. Leuchs, \textit{Lectures on Quantum Information} (Wiley-VCH,  2007).

\bibitem{qtech} K. Hornberger \etal, \rmp \textbf{84}, 157 (2012); J.-W. Pan, \etal, \rmp \textbf{84}, 777 (2012); M.~Poot, H.S.J.~van der Zant, Phys.~Rep. \textbf{511}, 273 (2012); P.~Shadbolt, \etal, Nature Phys. \textbf{10}, 278 (2014).

\bibitem{santos} E. Santos, Studies Hist. Phil. Science B \textbf{36}, 544 (2005).

\bibitem{overloop} A. Di Domenico, \etal, Found. Phys. \textbf{42}, 778 (2012); M. Giustina \etal, Nature \textbf{497}, 227 (2013); J. Gallicchio, A. S. Friedman, and D. I. Kaiser, \prl \textbf{112}, 110405 (2014).

\bibitem{wdc1} J.A.~Wheeler, in \cite{qf1}, p.~182;  A.J.~Leggett, in \cite{qf2}, p.~161; V.~Jacques, {\it et al.}, Science {\bf 315}, 966 (2007).

\bibitem{rmp} X. Ma, J. Kofler, and A. Zeilinger, {\em Delayed-choice gedanken experiments and their realizations}, arXiv:1407.2930 (2014).

\bibitem{hardy} L.~Hardy, {\em Reconstructing quantum theory}, arXiv:1303.1538.

\bibitem{ls} M.S.~Leifer and R.W.~Spekkens, \pra {\bf 88}, 052130 (2013).

\bibitem{chiribella} G.~Chiribella, G.M.~D'Ariano, and P.~Perinotti, \pra {\bf 84}, 012311 (2011).

\bibitem{fuchs} C.A.~Fuchs, {\em Notes on a Paulian Idea: Foundational, Historical, Anecdotal and Forward-Looking Thoughts on the Quantum}, arXiv:quant-ph/0105039.

\bibitem{pbr} M.F.~Pusey, J.~Barrett, and T.~Rudolph, Nature Phys.~{\bf 8}, 475 (2012).

\bibitem{cr2} R.~Colbeck and R.~Renner, \prl {\bf 108}, 150402 (2012); R.~Colbeck and R.~Renner, {\em A system's wave function is uniquely determined by its underlying physical state}, arXiv:1312.7353.

\bibitem{wdcq} R.~Ionicioiu and D.R.~Terno, \prl {\bf 107}, 230406 (2011)

\bibitem{wdc2} R.~Ionicioiu, T.~Jennewein, R.B.~Mann, and D.R.~Terno, Nature Comm. \textbf{5}, 4997  (2014).

\bibitem{gra86} P.~Grangier, G.~Roger, and A.~Aspect, Europhys. Lett.  \textbf{1},  173 (1986).

\bibitem{bohr} N.~Bohr, in P.A.~Schilpp (ed), \textit{Albert Einstein -- Philosopher-Scientist} (The Library of Living Philosophers, 1949), pp. 200-241; reprinted in \cite{qf1}, pp. 9--49.

\bibitem{six} L. C\'{e}leri, \etal, Found. Phys.  \textbf{44}, 576 (2014).

\bibitem{qdc-exp1} J.-S. Tang, \textit{et al.}, Nature Photonics {\bf 6}, 600 (2012), F. Kaiser,  \textit{et al.}, {Science} {\bf 338}, 637 (2012); A. Peruzzo, \textit{et al.}, Science {\bf 338}, 634 (2012);

\bibitem{qdc-exp2} S. Roy, A. Shukla, and T. S.  Mahesh, \pra {\bf 85}, 022109 (2012); R. Auccaise, \textit{et al.}, \pra {\bf 85}, 032121 (2012).

\bibitem{pt02} A.~Peres and D.R.~Terno, J. Mod. Opt. \textbf{49}, 1255 (2002).

\end{thebibliography}
\end{document}